\documentclass[12pt]{article}

\usepackage[normalem]{ulem}

\usepackage{xcolor}
\usepackage[english]{babel}
\usepackage[T1]{fontenc}
\usepackage[utf8]{inputenc}
\usepackage{authblk}
\usepackage{mathtools}
\usepackage{slashed}
\usepackage{amsmath,amssymb}
\usepackage{amsfonts}
\usepackage{graphicx,color,xcolor}
\usepackage{cite}
\usepackage{hyperref}
\hypersetup{
    colorlinks=true,
    linkcolor=blue,
    filecolor=magenta,      
    urlcolor=cyan,
}
\usepackage[capitalize]{cleveref}

\numberwithin{equation}{section} 

\definecolor{refcol}{rgb}{0.9,0.1,0.1}
\hypersetup{colorlinks=true,linkcolor=blue,citecolor=refcol,urlcolor=cyan,linktocpage}

\def\XXint#1#2#3{{\setbox0=\hbox{$#1{#2#3}{\int}$ }
\vcenter{\hbox{$#2#3$ }}\kern-.6\wd0}}

\newcommand{\ben}{\begin{eqnarray}\displaystyle}
\newcommand{\een}{\end{eqnarray}}

\newcommand{\be}{\begin{equation}}
\newcommand{\ee}{\end{equation}}

\newcommand{\bc}{\begin{center}}
\newcommand{\ec}{\end{center}}

\newcommand{\eesp}{\end{split}}
\newcommand{\bsp}{\begin{split}}

\newcommand{\Rmnum}[1]{\expandafter\@slowromancap\romannumeral #1@}

\newcommand{\e}{\epsilon}

\newcommand{\m}{\mu}								
\newcommand{\n}{\nu}								

\renewcommand{\r}{\rho}		

\newcommand{\x}{\xi}								

\newcommand{\I}{\infty}
\renewcommand{\L}{\Lambda}	

\newcommand{\cA}{\mathcal{A}}
\newcommand{\cB}{\mathcal{B}}
\newcommand{\cC}{\mathcal{C}}

\newcommand{\cF}{\mathcal{F}}

\newcommand{\cN}{\mathcal{N}}
\newcommand{\cO}{\mathcal{O}}

\newcommand{\cQ}{\mathcal{Q}}

\newcommand{\cS}{\mathcal{S}}

\newcommand{\cX}{\mathcal{X}}

\newcommand{\fS}{\mathfrak{S}}

\newcommand{\bensp}{\begin{eqnarray}\begin{split}}
\newcommand{\eensp}{\end{eqnarray}\end{split}}

\newcommand{\bnm}{\begin{matrix}}
\newcommand{\enm}{\end{matrix}}

\newcommand{\T}[{1}]{T^{#1}}

\def\XXint#1#2#3{{\setbox0=\hbox{$#1{#2#3}{\int}$ }
\vcenter{\hbox{$#2#3$ }}\kern-.6\wd0}}

\newcommand{\dow}{\partial}

\newcommand{\im}[1]{\text{Im}\left(#1\right)}
\newcommand{\re}[1]{\text{Re}\left(#1\right)}

\begin{document}
\begin{titlepage}
	\thispagestyle{empty}
	
	\title{
		{\Huge\bf  Freudenthal duality of near-extremal black holes and Jackiw-Teitelboim gravity }
	}
	
	\vfill
	
	\author{
		{\bf Arghya Chattopadhyay$^a$}\thanks{{\tt arghya.chattopadhyay@wits.ac.za}},  
		{\bf Taniya Mandal$^a$}\thanks{{\tt taniya.mandal@wits.ac.za}}\\    
		{\small{\it{$^a$National Institute of Theoretical and Computational Sciences,}\\ \it{School of Physics and Mandelstam Institute for Theoretical Physics,}\\ \it{University of the Witwatersrand, Wits, 2050, South Africa.}}}}
	
	\vfill
	
	\date{
		\vspace{1cm}
		\begin{quote}
			\centerline{{\bf Abstract}}
			{\small
				Freudenthal duality (F-duality), an anti-involution of charge vectors keep the entropy and attractor solutions invariant for an extremal supersymmetric black hole.  In this paper, we analyze the effect of F-duality on the entropy of a near-extremal $STU$ black hole in $\cN=2$ ungauged, four-dimensional supergravity. We consider double-extremal black holes, whose attractor solutions are fixed in terms of the black hole charges throughout the moduli space. It is well known that JT gravity governs the dynamics of the near-horizon regions of higher dimensional, near-extremal black holes. Owing to this fact, we reduce the four-dimensional supergravity theory to two dimensions to construct a Jackiw-Teitelboim (JT) gravity like model and compute the near-extremal entropy.  We then analyze the effect of F-duality on this entropy. We show that the F-duality breaks down for the case of near-extremal solutions if one considers the duality operation generated through near-extremal entropy rather than the extremal one.
			}
		\end{quote}
	}
	


\end{titlepage}
\thispagestyle{empty}\maketitle\vfill \eject

\tableofcontents

\section{Introduction}
Jackiw-Teitelboim (JT) gravity \cite{Teitelboim:1983ux, Jackiw:1984je},  a two-dimensional dilaton gravity model has recently become the pursuit of interest to many theoretical physicists, as it provides the simplest playground to study gravitational dynamics.  JT gravity, which is basically a particular class of two-dimensional dilaton gravity theories, does not have any propagating degrees of freedom. Thus the effective action crucially depends on the symmetries of the boundary theory. At the boundary of a JT gravity  in asymptotically $AdS_2$ spacetimes, the time reparametrization symmetry gets slightly broken. As a consequence, the solution to it is given by nearly $AdS_2$ or $NAdS_2$ spacetime that possess a nearly conformal or $NCFT_1$ symmetry \cite{Maldacena:2016upp}. Whereas, in the case of a nearly extremal black hole the near-horizon geometry is given by nearly $AdS_2$ ($NAdS_2$) geometry, which makes a manifest connection between JT gravity and near-extremal black hole\cite{Almheiri:2014cka, Maldacena:2016upp}. Similar to the nearly extremal black holes, SYK models \cite{Sachdev_1993, Kitaev:2017awl} have an approximate conformal symmetry ($NCFT_1$) in the infrared limit. This link prompts one to investigate the connection between JT gravity, near-extremal black holes and the SYK model at the low energy limit  \cite{Maldacena:2016hyu, Jensen:2016pah, Sarosi:2017ykf}. A suitable modification of JT gravity by higher derivative terms also captures the dynamics of higher derivative modified, higher-dimensional near-extremal black hole \cite{taniyacoming,Rathi:2021aaw}.

 More generally, JT gravity theories defined on any manifold with at least one asymptotic boundary depends on the choice of the boundary curves and one has to do a path integral over the different choices for these curves as well \cite{Maldacena:2016upp, Jensen:2016pah, Engelsoy:2016xyb}. These choices of boundary curves or \emph{wiggles} are governed by the same Schwarzian theory as the low energy limit of the SYK models. One can refer to \cite{Sarosi:2017ykf} for an excellent review and references therein to further explore this connection. This observation is further explored in \cite{Saad:2019lba} to show that JT gravity path integral on arbitrary orientable surfaces can be computed through ensembles of Hermitian matrices in a particular double scaling limit. This correspondence can also be generalised to different classes of matrix ensembles and JT gravity theories, even including JT supergravity theories \cite{Stanford:2019vob}. Relation between JT supergravity and matrix model and minimal strings also has been analysed including non-perturbative effects as well \cite{Johnson_2021, Johnson_2021_2}. JT model is also well studied in the context of $NAdS_2/NCFT_1$ holography \cite{Larsen:2018iou, Castro:2018ffi, Castro:2021wzn}.

In this paper, we are interested to study the behaviour of near-extremal black holes using JT gravity as a tool. Extremal black holes have a fascinating property of having finite entropy but zero temperature. An interesting phenomenon possessed by the extremal black hole is attractor mechanism \cite{Ferrara:1995ih, Ferrara:1997tw}, which says that the moduli fields always take the charge dependent fixed values at the horizon irrespective of their arbitrary asymptotic values. Thus the entropy of an extremal black hole always can be written in terms of the black hole charges only. There are numerous works on both supersymmetric and non-supersymmetric attractors \cite{Tripathy:2005qp, Dominic:2014zia} which has also been generalized for non-extremal black holes \cite{Goldstein:2014gta, Goldstein:2015pcl}. One can wonder whether the properties that hold for an extremal black hole in supergravity, can also hold true for a nearly extremal black hole as well. In particular, we are interested in studying the effect of the Freudenthal duality (F-duality) for near-extremal black holes. Freudenthal triple systems are long been known to mathematicians due to Hans Freudenthal \cite{FREUDENTHAL1954218} and then it has been connected to the entropy of a supersymmetric black hole \cite{Borsten:2009zy}. In the case of four-dimensional extremal, supersymmetric black holes, the entropy and corresponding attractor solutions are invariant under an anti-involutive operation of charges. This discrete duality is called Freudenthal duality \cite{FREUDENTHAL1954218, Borsten:2009zy, G_naydin_2010, Borsten_2013, Marrani_2013, Ferrara:2011gv, Ferrara_2013, marrani2015freudenthal, Fern_ndez_Melgarejo_2014, Borsten_2013_2,Klemm_2017}. It is different from the U-duality (electro-magnetic duality) as the quantities which remain invariant under U-duality change with Freudenthal duality except for entropy and attractor solution.

In \cite{Mandal:2017ioi},  it has been explicitly shown that for $\cN=2, IIA$ supergravity, an extremal supersymmetric $D0-D4-D6$ black hole is Freudenthal dual to an extremal supersymmetric $D0-D2-D4-D6$ black hole. Using H-FGK formalism \cite{Meessen:2011aa}, it has been shown in \cite{Galli:2012ji}, that the Freudnthal duality behaves as a part of a local symmetry for the case of both extremal and non-extremal black hole in $N=2, d=4$ ungauged supergravity. Freudenthal duality has also been studied in the context of $\cN>2$ supergravities \cite{Ferrara:2011gv}. In this paper, we consider $STU$ model in four-dimensional, ungauged, $\cN=2, IIA$ supergravity in asymptotically flat space, with all possible dyonic black hole charges. Entropy of the extremal, supersymmetric black hole solution in this model is Freudenthal invariant. Here we explicitly check, whether this duality is preserved in the case of a near-extremal black hole or not, in the context of JT-like theory that we get after dimensionally reducing the higher dimensional theory. Asymptotically flat, rotating, non-extremal black hole solutions have been studied in \cite{Chow:2013tia,Chow:2014cca,Sarosi:2015nja}. Our final aim is to study F-duality for these \emph{generic} non-extremal cases, this work is the initiation towards that by studying near-extremal versions.

In asymptotically flat space, the near-horizon region of a higher dimensional extremal (or near-extremal) black hole contains a throat like region with $AdS_2\times S^2$ geometry, whereas the far-horizon region is asymptotically flat. One of the solutions to the equations of motion that comes from the effective action\footnote{after integrating out the gauge fields \cite{Iliesiu:2020qvm}.} after dimensional reduction is  $AdS_2$ with a constant dilaton\footnote{Once we dimensionally reduce higher dimensional theory, a dilaton appears in lower-dimensional theory.} $\Phi_0$. We call it the background solution. It can be easily checked that, this background $AdS_2$ solution is the same as that appear in the near-horizon limit of the higher dimensional extremal black hole and $\Phi_0$ is the radius of both $AdS_2$ and $S^2$.  To go beyond extremality, we introduce a perturbation around this background solution. Expanding the reduced effective action around this fluctuation, we find a JT-like action. Following \cite{Maldacena:2016upp}, one can then find the entropy of the near-extremal black hole in terms of the value of the dilaton fluctuation at the boundary, namely $\phi_b$. As we are interested in the nature of higher dimensional physics, we fix the boundary at the overlapping region of near-horizon and far-zone geometry of higher dimensional black hole solution \cite{Nayak:2018qej, Iliesiu:2020qvm, Heydeman:2020hhw}. At this boundary, the dilaton fluctuation takes a fixed value, given by black hole charges. Thus effectively we can fix the entropy of a near-extremal black hole in terms of the black hole charges, where we considered that the charges remain fixed while moving away from extremality.

Semiclassically, at low temperature, the mass difference between  the extremal state and the near-extremal state $\Delta M=M-M_{ext}$ scales as $\Delta M\sim\frac{T^2}{M_{gap}}$, for a fixed charge black hole. Here $M$ and $T$ denote the mass and the temperature of the near-extremal state and $M_{ext}$ is the mass of the extremal state of the black hole. This questioned the reliability of the semiclassical analysis, as the black hole does not radiate at the temperature $T\lesssim M_{gap}$, since the required energy to Hawking radiate is of the order of the temperature $T$. Thus, earlier it was believed that there was a mass gap between the extremal state and the near-extremal state in the mass spectrum, at very low temperature when a fixed charge black hole was studied semiclassically. Including quantum effects, it has been shown in \cite{Iliesiu:2020qvm} that for a non-supersymmetric black hole mass gap can be removed, thus enabling black holes to Hawking radiate at any non zero temperature. As an effect, the entropy of a near-extremal black hole achieves an extra term, logarithmic in $(T \phi_b)$ in addition to a term linear in $T \phi_b$.  Whereas in \cite{Heydeman:2020hhw}, it has been shown that there is a mass gap for a near-extremal, near-BPS(near-supersymmetric) black hole solution in $\cN=2$ pure four-dimensional, ungauged supergravity by matching with $\cN=4$ super-JT gravity.   
In fact, the existence of mass gap is justified in this case as the degeneracy of the black hole is consistent with the Bekenstein-Hawking entropy formula \cite{Heydeman:2020hhw}. However in our current context, we are looking at the entropy of a near-extremal black hole in fermionic truncated supergravity, thus there is no mass gap in our black hole spectrum once we include quantum corrections.

Keeping the details for the main text, in a simplified form, F-duality for black holes can be understood as an anti-involution mapping of the charge vector $\cQ^\cA$ as 
\begin{equation}\label{eq:basicF}
F: \hat{\cQ}^\cA\rightarrow\Omega_{\cA\cB}{\dow S_0(\cQ)\over \dow \cQ^\cB}
\end{equation}
with $\Omega_{\cA\cB}$ being some symplectic metric and $ S_0(\cQ)$ is the extremal, supersymmetric entropy. Under this mapping $S_0(Q)$ remains invariant ($S_0(\hat{\cQ})=S_0(\cQ)$). We show that for double extremal attractor solution(for which the values of the moduli fields remain fixed in terms of the charges throughout the moduli space) for four-dimensional ungauged STU model the near extremal entropy takes the following form
\begin{align}
	S_{NE}=S_0+\mathcal{A}T S_0^{\frac{3}{2}}+\frac{3}{2}\log(\mathcal{B} T  S_0^{\frac{3}{2}}),
\end{align}
where $T$ is the temperature of the black hole. $\mathcal{A}$ and $\mathcal{B}$ are some charge and temperature independent constants. We analyse the effect of Freudenthal duality on this form of near-extremal entropy. We show that $S_{NE}$ is not invariant under F-duality if one takes the full entropy $S_{NE}$ as the function generating the F-dualization. Rather the F-duality invariance of black holes always have to be defined through the extremal entropy through \eqref{eq:basicF}, in which case $S_{NE}$ remains invariant.

We have structured the paper as the following. In \cref{sec:review}, we lay out a bare-bone detail of the bosonic sector of the $\cN=2$ supergravity and F-duality. We show the near-extremal entropy calculations in \cref{sec:JTlike}, following a dimensional reduction of the full four-dimensional metric to derive an effective two-dimensional dilaton gravity theory. In \cref{sec:nearlyF}, we discuss the fate of F-duality for the near extremal scenario and finally conclude in \cref{concu}. Further in \cref{app:STU}, we tabulate the necessary details of the STU black hole and the attractor solutions.


\section{Brief review of $\cN=2$ Supergravity and F-duality}\label{sec:review}
In four dimensions, the bosonic part of the $\cN=2$ supergravity action coupled with an arbitrary number of vector multiplets is given by
 \begin{equation}\label{eq:originallag}
S=\frac{-1}{8\pi G_4}\int d^4x \sqrt{-G}\Bigg(-{R\over 2}+h_{a\bar{b}}\dow_\mu x^a\dow_\nu x^{\bar{b}}G^{\mu\nu}-\mu_{\Lambda\Sigma}\cF^{\Lambda}_{\mu\nu}\cF^{\Sigma}_{\lambda\rho}G^{\mu\lambda}G^{\nu\rho}-\nu_{\Lambda\Sigma}\cF^{\Lambda}_{\mu\nu}*\cF^{\Sigma}_{\lambda\rho}G^{\mu\lambda}G^{\nu\rho}\Bigg)
\end{equation}
where $G_{\m\n}$ is the spacetime metric with the Ricci scalar $R$ and determinant $G$. $x^a$ are the $n$ complex moduli scalar fields with  the moduli space metric $h_{a\bar{b}}$. The field strength $\cF^\Lambda$ corresponds to the $(n+1)$ one form $A^\Lambda_\mu$, $\Lambda=0,1,\cdots,n$. The gauge coupling constants $\mu_{\Lambda\Sigma}$ and $\nu_{\Lambda\Sigma}$ and the moduli space metric are determined by the prepotential $F$. The number of the vector multiplets is denoted by $n$ and $G_4$ denotes the four-dimensional Newton's constant.

We are interested in a four-dimensional, $IIA$, $\cN=2$ supergravity that arises at the low energy limit when type $IIA$ string theory is compactified on a Calabi-Yau manifold $\mathcal{M}$, in the large volume limit. This theory is described by a holomorphic function called prepotential and that is given by
\begin{equation}\label{eq:theprepot}
	F=D_{abc}{X^aX^bX^c\over X^0},
\end{equation}
where the symplectic section $X^a$ is related to the complex moduli as  $ x^a={X^a\over X^0}$ . $D_{abc}$ is the triple intersection number  $D_{abc}=\frac{1}{6}\int_{\mathcal{M}} \alpha_a \wedge \alpha_b\wedge \alpha_c$ of two-form $ \alpha_a$, the basis of the cohomology group $H^2(\mathcal{M},Z)$. The Kahler potential is given by
\begin{equation}
K=-\log\left[i\sum_{\Lambda=0}^n \left(\overline{X^\Lambda}\dow_{\Lambda}F-X^\Lambda\overline{\dow_\Lambda F}\right)\right],
\end{equation}
which can be further simplified using the gauge $X^0=1$ as the following
\begin{equation}
	K=-\log\left[-i D_{abc}(x^a-\bar{x}^a)(x^b-\bar{x}^b)(x^c-\bar{x}^c)\right].
\end{equation}
The moduli space metric can be constructed as $h_{a\bar{b}}=\partial_a\partial_{\bar{b}}K$ i.e. by taking the partial derivative of the Kahler metric $K$ with respect to the complex moduli $ x^a$. The gauge coupling constants are given by the real and imaginary part of $\mathcal{N}$ as $\m=\im{\cN}$ and $\n=-\re{\cN}$, where
\begin{equation}\label{formN}
	\cN=\overline{F_{\Lambda\Sigma}}+2i{\im{F_{\Lambda\Omega}}\im{F_{\Pi\Sigma}}X^{\Omega}X^{\Pi}\over \im{F_{\Omega\Pi}}X^{\Omega}X^{\Pi}},
\end{equation}
and $F_{\Lambda\Sigma}=\dow_\Lambda\dow_\Sigma\, F$.

In this paper, for simplicity, we consider the STU model where the number of coupled vector multiplets is $n=3$ and only non zero intersection matrix is $D_{123}=1/6$. For $D0-D2-D4-D6$ black hole in STU supergravity, the entropy of an extremal, supersymmetric black hole is given by \cite{Shmakova:1996nz}
\begin{align}
S_{stu,susy}^{D0-D2-D4-D6}=\frac{\pi}{p^0}\sqrt{\frac{2}{3 d_{123}}\Delta_1\Delta_2\Delta_3-\left(p^0(p.q)-2D\right)^2},
\end{align}
where $\Delta_a=D_{abc}\tilde{x}^b\tilde{x}^c=3D_{abc}p^bp^c-p^0 q_a$ with real $\tilde{x}^a$ and $D=D_{abc}p^ap^bp^c$. Following \cite{Mandal:2017ioi}, one can write the above expression as 
\begin{equation}
 S_{stu,susy}^{D0-D2-D4-D6}={\pi\over 3p^0}\sqrt{{4\over 3}{(3D-p^0\,q_a p^a)^3\over D}-9(p^0\, (p.q)-2D)^2}.
 \end{equation}	
where $p.q=p^0q_0+p^a q_a$. $q_0$,$p^0$,$q_a$ and $p^a$ are the $D0$, $D6$, $D2$ and $D4$ brane charges. As mentioned in \cite{Mandal:2015mke}, for a generic supergravity model, there could be more than one sets of supersymmetric attractor solution with different form of entropies for different charge sectors depending on the value of an involutory matrix $I^a_b$. The properties of $I^a_b$ are $$ I^a_cI^c_b=\delta^a_b, \quad \text{and} \quad D_{ade}I^d_b I^e_c=D_{abc}.$$ For STU model, there is only one choice of the matrix where $I^a_b=\delta^a_b$. Thus in STU model we have only one expression for the entropy for all possible D-brane charges.

Now we will briefly summarise F-duality, which is a non-linear, anti-involutive transformation of charge vector $\cQ^M=(p^\Lambda,q_\Lambda)$ such that the entropy and the attractor values of an extremal, supersymmetric black hole remain fixed under this transformation. The transformation acts as follows
\begin{align}
& \pi\hat{\cQ}^M=\Omega^{MN}\frac{\partial S(\cQ)}{\partial \cQ^N},\nonumber\\
& \hat{\hat{\cQ}}=-\cQ, \quad \text{and} \quad S(\hat{\cQ})=S(\cQ) 
\end{align}
where $S(\cQ)$ is the extremal entropy of a supersymmetric black hole corresponding to the charge vector $\cQ^M=(p^\Lambda,q_\Lambda)$. $\Omega^{MN}$ is $2(n+1)\times2(n+1)$ symplectic matrix with $\Omega^T=-\Omega$ and $\Omega^2=-I$.  We consider $\Omega=\begin{pmatrix}
	0 &-I\\
	I& 0
\end{pmatrix}$.
   
Thus by F-dualizing an extremal $D0-D2-D4-D6$ supersymmetric black hole, one would expect to get another extremal $D0-D2-D4-D6$ supersymmetric black hole whose charges non-linearly depend on that of the former one. Nevertheless, an extremal $D0-D2-D4-D6$ supersymmetric black hole can also be mapped to an extremal $D0-D4-D6$ supersymmetric black hole via F-duality \cite{Mandal:2017ioi}.

\section{JT-like theory from $D=4$, $\cN=2$ Supergravity}\label{sec:JTlike}
Extremal black holes are interesting objects both in classical gravity and supergravity by their own virtue. In both the cases, near-horizon geometry of a four-dimensional extremal, charged black hole in asymptotically flat space, factorises as $AdS_2\times S^2$ with same radius for $AdS_2$ and $S_2$. In this section, we show that by starting with a supergravity theory in four dimensions, we find a JT-like dilaton gravity theory in two dimensions upon dimensional reduction. This JT-like theory captures the dynamics of the near-extremal black hole solution of the supergravity theory, where the extremal solution that we are interested in is supersymmetric. It has been seen that, for an extremal black hole in supergravity, irrespective of whether it is supersymmetric or not, the entropy is fixed in terms of black hole charges. Also, the values of the moduli fields are fixed in terms of charges at the horizon, we call them attractor solutions. 
We are interested in the fate of F-duality once we perturb away from the extremal scenario.

\subsection{Dimensional Reduction}
We start with a dimensional reduction on the sphere at the near-horizon region by considering a spherically symmetric ansatz for the metric as \footnote{This is not the most generic ansatz for dimensional reduction. Considering the most generic ansatz, in lower dimensions, one gets Maxwell-Dilaton gravity theory with $SO(3)$ Yang-Mills field\cite{Iliesiu:2020qvm}. Integrating out the Yang-Mills field, one finds a rotational solution with angular momentum $j$\cite{Iliesiu:2020qvm}. As our goal is to comment on Freudenthal duality of a spherically symmetric, non-rotating, near-extremal black hole, we can set $j=0$. This is equivalent to considering the ansatz \eqref{dimredanst} for dimensional reduction.}\textsuperscript{,}\footnote{We are considering $G_{\m\n}$ to be asymptotically flat.}
\begin{equation}\label{dimredanst}
	ds^2=G_{\mu\nu}dx^\mu dx^\nu=\tilde{g}_{\alpha\beta}dx^{\alpha}dx^{\beta}+{\Phi}^2 d\Omega_2^2.
\end{equation}
We are interested in dyonic black hole solutions i.e having both electric charges $Q_\Lambda$ and magnetic charges  $P^\Lambda$. Dimensionally reducing the bulk action over \eqref{dimredanst} with further assuming that $ \cF^\Lambda_{\theta\phi}=P^{\Lambda}\sin\theta$, we find 
\begin{align}
\tilde{S}_{bulk}&=-8\pi G_4 S_{bulk}=4\pi\int d^2x \sqrt{-\tilde{g}}\, \Phi^2 \Bigg[-{R(\tilde{g})\over 2}-{1\over \Phi^2}(\nabla_\alpha\Phi)^2-\frac{1}{\Phi^2}-\mu_{\Lambda\Sigma}\cF^\Lambda_{\alpha\beta} \cF^{\Sigma\,\alpha\beta}\nonumber\\
&-\frac{2}{\Phi^4}\mu_{\Lambda\Sigma}P^\Lambda P^\Sigma\Bigg]-4\pi\int d^2x (4\nu_{\Lambda\Sigma}P^\Lambda \cF^\Sigma_{rt})+8\pi\int d^2x  \sqrt{\tilde{g}}\,  \nabla_\alpha(\Phi\nabla^\alpha\Phi)\nonumber\\
&+4\pi\int d^2x\,\sqrt{-\tilde{g}}\,\Phi^2\,h_{a\bar{b}}\dow_\alpha x^a\dow_\beta x^{\bar{b}}\tilde{g}^{\alpha\beta}.
\end{align}
One should note that the scalar fields $x^a$ only have a dependence on the radial direction. For our convenience, we write this reduced action in the following way by introducing a two-dimensional Levi-Civita symbol $\epsilon^{\alpha\beta}$ such that  $\epsilon^{rt}=1$,
\begin{align}
&\tilde{S}_{bulk} =4\pi\int d^2x \sqrt{-\tilde{g}}\, \Phi^2 \left[-{R(\tilde{g})\over 2}-{1\over \Phi^2}(\nabla_\alpha\Phi)^2-\frac{1}{\Phi^2}-\mu_{\Lambda\Sigma}\cF^\Lambda_{\alpha\beta} \cF^{\Sigma\,\alpha\beta}-\frac{2}{\Phi^4}\mu_{\Lambda\Sigma}P^\Lambda P^\Sigma\right]\nonumber\\
&-4\pi\int d^2x (2\nu_{\Lambda\Sigma}P^\Lambda \epsilon^{\alpha\beta}\cF^\Sigma_{\alpha\beta})+8\pi\int dr\,dt  \sqrt{\tilde{g}}\,  \nabla_\alpha(\Phi\nabla^\alpha\Phi)+4\pi\int d^2x\,\sqrt{-\tilde{g}}\,\Phi^2\,h_{a\bar{b}}\dow_\alpha x^a\dow_\beta x^{\bar{b}}\tilde{g}^{\alpha\beta}.
\end{align}
This two-dimensional action contains a derivative on the dilation $\Phi$. As our target is to get a JT-like theory, we do not want a term like that in the two-dimensional action. A simple way to get rid of that is Weyl rescaling of the metric in the following way
\begin{equation}
	g_{\alpha\beta}={\Phi\over \Phi_0}\tilde{g}_{\alpha\beta}.
\end{equation}
%
The Weyl rescaled bulk action is then given by 
\begin{align}
\tilde{\cS}_{bulk} &= 4\pi\int d^2x \sqrt{-g}\, \Phi^2 \left[-{R(g)\over 2}-{\Phi_0\over\Phi^3}-{\Phi\over\Phi_0}\mu_{\Lambda\Sigma}\cF^\Lambda_{\alpha\beta} \cF^{\Sigma\,\alpha\beta}-\frac{2\Phi_0}{\Phi^5}\mu_{\Lambda\Sigma}P^\Lambda P^\Sigma\right]\nonumber\\
&-4\pi\int d^2x (2\nu_{\Lambda\Sigma}P^\Lambda\cF^\Sigma_{\alpha\beta}\epsilon^{\alpha\beta})+6\pi\int d^2x  \sqrt{-g}\,  \nabla_\alpha(\Phi\nabla^\alpha\Phi)+4\pi\int d^2x \,\sqrt{-g}\,\Phi^2\,h_{a\bar{b}}\dow_\alpha x^a\dow_\beta x^{\bar{b}}g^{\alpha\beta}.
\end{align}
%
We also need to dimensionally reduce the Gibbons-Hawking-York(GHY) boundary term and the boundary term for the gauge field as well. From \eqref{eq:originallag}, one can check that the three-dimensional GHY boundary term corresponding to $-8\pi G_4 S$ is given by
\begin{equation}
	S^{(3)}_{\text{GHY}}=-\int d^3x\,\sqrt{H}\, K^{(3)},
\end{equation}
where $H$ is the induced hypersurface metric of the full metric $G$ and $K^{(3)}$ is the three-dimensional extrinsic curvature tensor. Defining $h_{\mu\nu}$ as the induced metric and $\hat{K}$ as the reduced extrinsic curvature from the reduced $1D$ point of view one can write
\begin{equation}
S_{\text{GHY}}=-4\pi\int \sqrt{-h}\,\Phi^2\left(\hat{K}+{2\over \Phi}\hat{n}.\nabla\Phi\right),
\end{equation}
where $\hat{n}_\alpha$ is the unit normal vector on the hypersurface. After Weyl rescaling, this can be re-written as 
\begin{equation}
	\cS_{\text{GHY}}=-4\pi\int\sqrt{-h}\,\Phi^2\left(K+{3\over 2\Phi}n.\nabla\Phi\right),
\end{equation} 
where $n$ is the Weyl rescaled unit vector and $K$ is the corresponding extrinsic curvature. Three-dimensional boundary term needed for the well behaved variation of $-8\pi G_4 S$ with respect to the gauge field is  given by\footnote{As we are interested in fixed charge black hole solution, we need to add a boundary term for the gauge field as \eqref{gaugeboundary}\cite{Chamblin:1999tk}. }
\begin{equation}\label{gaugeboundary}
	S^{(3)}_{gauge}=4\int d^3x\,\sqrt{-H}\left[\mu_{\Lambda\Sigma}\,n_\mu\cF^{\Lambda \mu\nu}A^\Sigma_{\nu}+\nu_{\Lambda\Sigma}\,n_\mu*\cF^{\Lambda \mu\nu}A^\Sigma_{\nu}\right].
\end{equation}
After dimensional reduction and the Weyl transformation, this term can be written as 
\begin{equation}
	\cS_{gauge}=16\pi \int dt\,\sqrt{-h}\,{\Phi^3\over \Phi_0}\mu_{\Lambda\Sigma}\,n_\alpha \cF^{\Lambda\,\alpha\beta}A_{\beta}^\Sigma+16\pi\int dt\,{\sqrt{-h}\over \sqrt{-g}}\,\nu_{\Lambda\Sigma}{\Phi^2\over\Phi_0^2}\,n_r P^{\Lambda}A_{t}^\Sigma
\end{equation}
Thus the full dimensionally reduced, Weyl rescaled action is
\begin{align}\label{eq:fulldimred}
	&\cS_{tot}\
	=-4\pi\int d^2x\sqrt{-g}\,  \left[{\Phi^2R(g)\over 2}+{\Phi_0\over\Phi}+{\Phi^3\over\Phi_0}\mu_{\Lambda\Sigma}\cF^\Lambda_{\alpha\beta} \cF^{\Sigma\,\alpha\beta}+\frac{2\Phi_0}{\Phi^3}\mu_{\Lambda\Sigma}P^\Lambda P^\Sigma-\Phi^2\,h_{a\bar{b}}\dow_\alpha x^a\dow_\beta x^{\bar{b}}g^{\alpha\beta}\right]\nonumber\\
	&-4\pi\int d^2x (2\nu_{\Lambda\Sigma}P^\Lambda\cF^\Sigma_{\alpha_\beta}\epsilon^{\alpha^\beta})+16\pi \int dt\,\sqrt{-h}\left[{\Phi^3\over \Phi_0}\mu_{\Lambda\Sigma}\,n_\alpha \cF^{\Lambda\,\alpha\beta}A_{\beta}^\Sigma+{1\over \sqrt{-g}}\,\nu_{\Lambda\Sigma}\,n_r P^{\Lambda}A_{t}^\Sigma\right]\nonumber\\
	&-4\pi\int\sqrt{-h}\,\Phi^2\,k.
\end{align}
Corresponding dilaton and gravity equations of motion are
 \begin{align}\label{eq:reduceom}
 	&\Phi R(g)-{\Phi_0\over \Phi^2}+3{\Phi^2\over \Phi_0}\mu_{\Lambda\Sigma}\cF^{\Lambda}_{\alpha\beta}\cF^{\Sigma\,\alpha\beta}-6{\Phi_0\over \Phi^4}\mu_{\Lambda\Sigma}P^\Lambda P^\Sigma-2\Phi h_{a\bar{b}}\dow_\alpha x^a\dow_\beta x^{\bar{b}}g^{\alpha\beta}=0
 	\end{align}
 	\begin{align}\label{eq:reduceom11}
 &-{1\over 2}g_{\gamma\delta}\left[-{\Phi^2 R\over 2}-{\Phi_0\over \Phi}-{\Phi^3\over \Phi_0}\mu_{\Lambda\Sigma}\cF^\Lambda_{\alpha\beta}\cF^{\Sigma\,\alpha\beta}-{2\Phi_0\over \Phi^3}\mu_{\Lambda\Sigma}P^\Lambda P^\Sigma+\Phi^2h_{a\bar{b}}\dow_\alpha x^a\dow_\beta x^{\bar{b}}g^{\alpha\beta}\right]-{\Phi^2\over 2}R_{\gamma\delta}\nonumber\\
 &-{2\Phi^3\over \Phi_0}\mu_{\Lambda\Sigma}\cF^\Lambda_{\delta\beta}\cF^{\Sigma}_{\gamma\psi}g^{\beta\psi}+\Phi^2h_{a\bar{b}}\dow_\gamma x^a\dow_\delta x^{\bar{b}}-\nabla_\gamma\Phi\nabla_\delta\Phi-\Phi\nabla_\gamma\nabla_\delta\Phi+\left((\nabla^\mu\Phi)^2+\Phi\nabla^2\Phi\right)g_{\gamma\delta}=0
 \end{align}
We also have equations for gauge fields and the moduli fields. As our case of interest is the double extremal  solution i.e. when the moduli fields take constant value in terms of the black hole charges throughout the moduli space, equations corresponding to them are trivially satisfied.  Consequently \eqref{eq:reduceom} and trace of \eqref{eq:reduceom11} simplifies to
\begin{align}\label{eq: dilatoneq}
	&\Phi R(g)-{\Phi_0\over \Phi^2}+3{\Phi^2\over \Phi_0}\mu_{\Lambda\Sigma}\cF^{\Lambda}_{\alpha\beta}\cF^{\Sigma\,\alpha\beta}-6{\Phi_0\over \Phi^4}\mu_{\Lambda\Sigma}P^\Lambda P^\Sigma=0\\
	&{\Phi_0\over \Phi}-{\Phi^3\over \Phi_0}\mu_{\Lambda\Sigma}\cF^{\Lambda}_{\alpha\beta}\cF^{\Sigma\,\alpha\beta}+{2\Phi_0\over\Phi^3}\mu_{\Lambda\Sigma}P^\Lambda P^\Sigma+(\nabla_\mu \Phi)^2+\Phi\nabla^2\Phi=0.
\end{align}
Remembering that the gauge field equation of motion coming out of the full $4$D theory has the property in the near-horizon limit 
\begin{align}
\partial_\mu\left(\sqrt{g}\Phi^2 \mu_{\Lambda_\Sigma}\mathcal{F}^{\Sigma\mu\nu}+\nu_{\Lambda_\Sigma}P^\Sigma \epsilon^{\mu\nu}\right)&=0,
\end{align}
the solution to the above equation can be written as
\begin{align}\label{4sol}
\sqrt{g}\Phi^2 \mu_{\Lambda_\Sigma}\mathcal{F}^{\Sigma\mu\nu}+\nu_{\Lambda_\Sigma}P^\Sigma \epsilon^{\mu\nu}&=-Q_\Lambda \epsilon^{\mu\nu},
\end{align}
with the usual convention $\epsilon^{rt}=-\epsilon^{tr}=1$ and $\epsilon_{rt}=g\,\epsilon^{rt}$.  For convenience, we rescale the electric and magnetic charges as $P^\Lambda=\sqrt{\frac{G_4}{2}}p^\Lambda$ and $Q_\Lambda=\sqrt{\frac{G_4}{2}}q_\Lambda$. Then in two dimension, following \eqref{4sol}, the solution to the gauge field equation takes the following form
\begin{align}\label{2sol}
\sqrt{g}\frac{\Phi^3}{\Phi_0} \mu_{\Lambda_\Sigma}\mathcal{F}^{\Sigma\mu\nu}+
\sqrt{\frac{G_4}{2}}\nu_{\Lambda_\Sigma}p^\Sigma \epsilon^{\mu\nu}&=-\sqrt{\frac{G_4}{2}}q_\Lambda \epsilon^{\mu\nu}.
\end{align}
To get a JT-like action, we integrating out the gauge fields from \eqref{eq:fulldimred} .  The effective theory is given by the following action
\begin{align}\label{2deffac}
	\fS=&4\pi\int d^2x\,\sqrt{-g}\left[-{\Phi^2R\over 2}-{\Phi_0\over \Phi}-{2\Phi_0\over \Phi^3}\left(\cX+\frac{G_4}{2}\mu_{\Lambda\Sigma}p^\Lambda p^\Sigma\right)\right]-4\pi\int dt\sqrt{-h}\Phi^2 k.
\end{align}
where 
\begin{align}
 \cX(\mu_{\Lambda,\Sigma},\nu_{\Lambda,\Sigma},P,Q)=\frac{G_4}{2}\Bigg((\mu^{-1})^{\Lambda\Sigma}q_\Lambda q_\Sigma+p^\Sigma {(\nu\mu^{-1})_\Sigma}^\Lambda q_\Lambda+q_\Lambda {(\mu^{-1}\nu)^\Lambda}_\Sigma p^\Sigma+p^\Lambda (\nu\mu^{-1}\nu)_{\Lambda\Sigma}p^\Sigma\Bigg).
\end{align}
We can also get \eqref{2deffac}, using the on-shell value \eqref{2sol} of the gauge field strength in \eqref{eq:fulldimred}. The equations of motion derived from this effective action are
\begin{equation}\label{eq:2deom}
	\begin{split}
	\Phi R(g)-{\Phi_0\over \Phi^2}-6{\Phi_0\over \Phi^4}\left(\cX+\frac{G_4}{2}\mu_{\Lambda\Sigma}p^\Lambda p^\Sigma\right)&=0\\
1+{2\over\Phi^2}\left(\cX+\frac{G_4}{2}\mu_{\Lambda\Sigma}p^\Lambda p^\Sigma\right)+{\Phi\over \Phi_0}(\nabla_\mu \Phi)^2+{\Phi^2\over \Phi_0}\nabla^2\Phi&=0.
\end{split}
\end{equation}
This alternate action can be simply viewed as some $2D$ dilaton theory as $\fS=\fS_{bulk}+\fS_{bdy}$, where
\begin{equation} \label{eq:altenrate2d}
	\begin{split}
		\fS_{bulk}&=-{1\over 2 G_4}\int d^2x\,\sqrt{-g}\left[-{\Phi^2R\over 2}-U(\Phi)\right]\\
		\fS_{bdy}&={1\over 2 G_4}\int dt\sqrt{-h}\Phi^2 K.
		\end{split}
\end{equation}
with the effective dilaton potential $\displaystyle{U(\Phi)={\Phi_0\over \Phi}+{2\Phi_0\over \Phi^3}\left(\cX+\frac{G_4}{2}\mu_{\Lambda\Sigma}p^\Lambda p^\Sigma\right)}$. We have inserted the normalisation $\frac{-1}{ 8\pi G_4}$ in front of the action to carefully track down the gravitational constant now onwards.

One of the solution to \eqref{eq:2deom} is given by
\begin{align}\label{solution2d}
\Phi &= r, \quad \text{with} \quad 
ds^2 = \frac{\Phi}{\Phi_0} \left(-f(r) dt^2 +\frac{dr^2}{f(r)}\right)\nonumber\\
 f(r)& = 1+\frac{\mathcal{C}}{\Phi}+\frac{2\Theta(\cQ)}{\Phi^2},\quad \Theta(\cQ)=-\left(\cX+\frac{G_4}{2}\mu_{\Lambda\Sigma}p^\Lambda p^\Sigma\right).
\end{align}
where $\mathcal{C}$ is an integration constant, which we fix shortly.  As elaborately explained in \cite{Nayak:2018qej} and furthered in \cite{Iliesiu:2020qvm, Heydeman:2020hhw} it is always beneficial to consider the near horizon and the far horizon regions of the four-dimensional, near extremal black hole spacetime separately. Near horizon region(NH) can be well approximated as $AdS_2\times S^2$. This approximation holds true in the region where the radial distance
$
	{r-r_0}\ll r_0,
$
with $r_0$ being the horizon radius for the extremal case. The far horizon(FH) metric remains well approximated by the extremal case even in the non-extremal scenario and it is given by 
\begin{equation}\label{44sol}
	ds^2=-f(r) dt^2+{dr^2\over f(r)}+r^2 d\Omega_2^2.
\end{equation}
In the extremal limit, near the horizon, the metric \eqref{44sol} reads as
\begin{equation}
 \quad ds^2=-\frac{(r-r_0)^2}{L_2^2}dt^2+\frac{L_2^2}{(r-r_0)^2}dr^2+r_0^2 d\Omega_2^2 , \quad L_2=r_0,
\end{equation}
whereas the solution \eqref{solution2d} takes the following form 
\begin{equation}\label{soln2dd}
 \quad ds^2=-\frac{(r-r_0)^2}{L_2^2}dt^2+\frac{L_2^2}{(r-r_0)^2}dr^2 , \quad \Phi=r_0.
\end{equation}
Using this blackening factor as the boundary condition one can now fix the integration constant in \eqref{solution2d} as $\cC=-2r_0$ in the large $r$ limit. Also in this limit, $\Theta(Q)=\frac{r_0}{2}$. Both of these regions overlap inside the bulk where the radial distance satisfies the condition that the $AdS_2$ radius $L_2\ll r-r_0\ll r_0$. One should note at this point that in the present context we are dealing with an asymptotically flat four-dimensional metric and therefore for all our purposes $L_2=r_0=\Phi_0$ for the extremal situation. Interestingly, if one considers a holographic screen at an arbitrary radial distance $r=r_0+r_b$ inside the overlapping region of NH and FH, the leading low temperature behaviour is independent of the choice of $r_b$ \cite{Nayak:2018qej,Iliesiu:2020qvm}. In the following, we will use this crucial aspect to fix the near extremal entropy as a function of the dyonic charges.

\subsection{Near Extremal Entropy}

So far in our discussions, we only focused on the near horizon region and dimensionally reduced the action corresponding to \eqref{eq:originallag} on $S^2$ . We also derived the two-dimensional, effective dilaton gravity action \eqref{eq:altenrate2d}. At the horizon, dilaton $\Phi$ takes the constant value $r_0$ which gets fixed from \eqref{eq:2deom} as
\begin{equation}\label{eq:dilatonatheorizon}
	r_0^2=2\Theta(\cQ),\quad R=-{2\over \Phi_0^2}.
\end{equation}
In the near-horizon region, where $r-r_0 \ll r_0$, the value of the dilaton and the metric are given by small perturbation around the background solution i.e. constant dilaton and $AdS_2$ metric. In two dimensions, the metric can be written in terms of a single independent parameter, in conformal gauge we write the metric as
\begin{align}\label{gensol}
	ds^2=e^{2\omega}(-dt^2+d\rho^2).
\end{align}
 For the background solution we set $\omega=\omega_0$. Using the coordinate transformation 
 \begin{equation}\label{gp}
\frac{{L_2}^2}{\r}=r-r_0
 \end{equation}
 where $r_0$ is the horizon of extremal solution and identifying $L_2=r_0$  one can write the two-dimensional background metric \eqref{soln2dd} with constant dilaton as
\begin{align}
	ds^2=e^{2\omega_0}(-dt^2+d\rho^2), \quad e^{2\omega_0}={L_2^2\over \rho^2}
\end{align}
where $\rho\equiv\rho(r)$ is a function of the original radial coordinate. For \eqref{gensol} the Ricci scalar simply boils down to
\begin{equation}
	R=-2e^{-2 \omega}\left({\dow^2\omega\over \dow \rho^2}-{\dow^2\omega\over \dow t^2}\right)
\end{equation}
Therefore the bulk part of the dilaton action \eqref{eq:altenrate2d} can be written for  \eqref{gensol} and a generic dilaton $\Phi$ as 
\begin{equation}
	\fS_{bulk}={1\over 2 G_4}\int d^2x\, \sqrt{-g}\left[\Phi^2 e^{-2 \omega}\left({\dow^2\omega\over \dow \rho^2}-{\dow^2\omega\over \dow t^2}\right)-U(\Phi)\right].
\end{equation}
The effective potential $U(\Phi)$ vanishes for the background constant dilaton solution \eqref{soln2dd}. To evaluate the full action \eqref{eq:altenrate2d} one can now switch to the Euclidean time $\tau=it$ and simply write the extrinsic curvature scalar $K={1/r_0}$. 
 The full action \eqref{eq:altenrate2d} for this background solution, provides the extremal entropy. Using attractor functional method \cite{Sen:2007qy}, it can be seen that  the extremal entropy is
\begin{align}\label{eq:s0}
	S_{0}=\frac{\pi r_0^2}{G_4}= {2\pi\lvert\Theta(\cQ)\rvert\over G_4} .
\end{align}
In terms of charges, it takes the explicit form as bellow \cite{Ferrara:1997tw},
\begin{align}
S_{0} &= \frac{\pi}{2}\lvert (\mu^{-1})^{\Lambda\Sigma}q_\Lambda q_\Sigma+p^\Sigma {(\nu\mu^{-1})_\Sigma}^\Lambda q_\Lambda+q_\Lambda {(\mu^{-1}\nu)^\Lambda}_\Sigma p^\Sigma\nonumber\\&+p^\Lambda (\nu\mu^{-1}\nu)_{\Lambda\Sigma}p^\Sigma +\mu_{\Lambda\Sigma}p^\Lambda p^\Sigma\rvert_{horizon}
\end{align}
In the above formula, the values of $\mu_{\Lambda\Sigma}$ and $\nu_{\Lambda\Sigma}$ have to be evaluated at the horizon using attractor values and \eqref{formN}. For a STU black hole with all possible D brane charges in $IIA$ supergravity, one can use the attractor values as pointed out in \cref{app:STU} and rewrite the entropy as  \cite{Shmakova:1996nz},
\begin{eqnarray}\label{eq:entropystu}
	S_{stu}^{D0-D2-D4-D6}=\frac{\pi}{p^0}\sqrt{\frac{2}{3}\Delta_1\Delta_2\Delta_3-\left(p^0(p.q)-2D\right)^2}.
\end{eqnarray}
 It has been shown in \cite{Mandal:2017ioi} that $D0-D2-D4-D6$ black hole has a Freudenthal dual black with charge configuration $D0-D4-D6$. One can further show that a STU $D0-D2-D4-D6$ black hole can also be F-dual to another $D0-D2-D4-D6$ black hole.

We will now \emph{switch-on} the perturbation around the dilaton and the metric as 
\begin{align}\label{fluc}
	\Phi=r_0(1+\phi)\quad, \omega=\omega_0+ \Omega.
\end{align}
 With this scheme of perturbation $\fS$ can be written as $\fS_{bulk}=\fS_{bulk}^0+ \fS_{bulk}^1+ \,\fS_{bulk}^2+\cO(\phi^3)$, where
\begin{equation}
	\begin{split}
	\fS_{bulk}^0&={1\over 2 G_4}\int d^2x\, \sqrt{-g}\left[r_0^2 e^{-2 \omega}\left({\dow^2\omega\over \dow \rho^2}-{\dow^2\omega\over \dow t^2}\right)-U(\Phi_0)\right]\\
	\fS_{bulk}^1 &=\frac{\Phi_0^2}{2G_4}\int\sqrt{g}\left(- \phi(t,z)\left(-\frac{R}{2}-\frac{1}{\Phi_0^2}\right)\right)
	\end{split}
\end{equation}
The boundary action can also be written as a power series of $\phi$ and $\Omega$ as $\fS_{bdry}=\fS_{bdry}^0+ \fS_{bdry}^1+ \,\fS_{bdry}^2+\cO(\phi^3)$. As said earlier, $\fS_{bulk}^0+\fS_{bdry}^0$ provides the extremal entropy. Using \eqref{eq:dilatonatheorizon}, we see that $\fS_{bulk}^1$  vanishes and we are left with the boundary action  $\fS_{bdry}^1$ linearly in $\phi$. In Euclidean signature, the boundary action is  as
\begin{align}\label{eq:ghyaction}
\fS_{bdry}^1=-\frac{r_0^2}{G_4}\int dx\, \sqrt{h} \phi K
\end{align}
To evaluate the above integral, we consider the curve separating the near horizon and far horizon region of four-dimensional, near extremal black hole as our boundary. On this boundary, the induced metric is $h=\frac{L_2^2}{\epsilon^2}$ and proper length is $\int du \sqrt{h} = \frac{\beta L_2}{\epsilon}$. Here, $\beta$ is the periodicity of the boundary time $u$ and the constant boundary value of the dilaton is  $\phi\lvert_{bdry}=\frac{\phi_b}{\epsilon}$. This boundary cuts out a patch of near horizon $AdS_2$.  In Poincare coordinate, $AdS_2$ metric takes the form $ds^2=\frac{L_2^2}{\rho^2}(dt^2+d\rho^2)$ which implies $g_{uu}=\frac{L_2^2}{\rho^2}(t'^2+\rho'^2)=\frac{L_2^2}{\epsilon^2}$. Solution to this equation gives the equation of boundary curve as $\rho(u)\simeq\epsilon t'(u)$ in the small $\epsilon$ limit, where $'$ denotes the derivative with respect to the boundary time $u$. Then the extrinsic curvature at the boundary takes the following form
\begin{eqnarray}
		K={t'(t'^2+\rho'^2+\rho \rho'')-\rho \rho't''\over {L_2}(t'^2+\rho'^2)^{3\over2}}={1\over L_2}(1+\e^2 \,\text{sch}(t,u))
	\end{eqnarray}
where
 \begin{equation}
	\text{sch}(t,u)=-{1\over 2}\left({t''\over t'}\right)^2+\left({t''\over t'}\right)'.
\end{equation}
Thus the boundary action can be written as 
\begin{equation}\label{eq:naives1}
	\fS_{bdry}^1 =-{r_0^2\over{G_4}}\int_0^\beta du\, \phi_b\, \text{sch}(t,u)
\end{equation}
Corresponding  partition function that includes quantum corrections, takes in to account the path integrals over all possible boundaries $t(u)$ up to $SL(2,R)$ identification and is given by \cite{Iliesiu:2020qvm, taniyacoming},
\begin{eqnarray}
	Z_{\text{sch}}(\phi_b,\beta) &=& \left(\frac{\hat{\phi}_b}{\beta}\right)^{\frac{3}{2}}e^{\frac{2\pi^2 \hat{\phi}_b}{\beta}}
\end{eqnarray}
where $\hat{\phi_b}={r_0^2\over{G_4}}\phi_b$. Entropy derived from this partition function provides the entropy of near extremal black hole above extremality and is \cite{Iliesiu:2020qvm},
\begin{eqnarray}\label{exen}
\delta S =\frac{3}{2}+4\pi^2\hat{\phi_b} T+\frac{3}{2} \log(T\hat{\phi_b})
\end{eqnarray}
where we use the identification of the temperature $T$ with the inverse of the periodicity of boundary time as $T=1/\beta$. Now we need to interpret the value of $\phi_b$ in terms of the parameter of the near extremal black hole.

The generic solution for the equation of dilaton derived from \eqref{eq:2deom} is given by $\Phi =r$, where at the horizon of the extremal solution it takes the value $\Phi =r_0$. Thus at any point near the horizon it takes the value given by $\Phi=r_0(1+\phi)$ as \eqref{fluc}. In global coordinate, we fix the position of the boundary, separating NH and FH as $r=r_0+r_b$. Dilaton takes the constant value $\Phi=r=r_0(1+\frac{\phi_b}{\epsilon})$ at the boundary. Thus at the boundary $\phi_b=\frac{(r-r_0)\epsilon}{r_0}$. Using the relation between global coordinate and Poincaré patch \eqref{gp}, one can see  that $\phi_b=\frac{L_2^2}{r_0}=r_0$ at the boundary where $\rho\rightarrow\epsilon$ \cite{Iliesiu:2020qvm}. Thus we find from \eqref{exen}
\begin{eqnarray}\label{exennew}
	\delta S =\frac{3}{2}+\frac{4\pi^2}{G_4} r_0^3 T+\frac{3}{2} \log\left(\frac{r_0^3T}{G_4}\right).
\end{eqnarray}
The near extremal entropy is 
\begin{align}\label{eq:fullS}
S_{NE}=S_0+\delta S =\frac{\pi r_0^2}{G_4}\left(1+4\pi T r_0\right)+\frac{3}{2} \log\left(\frac{r_0^3T}{G_4}\right)+  \frac{3}{2},
\end{align}
which is now manifestly independent of the boundary.

\section{Near extremal limit and F-duality}\label{sec:nearlyF}
At this point it is worth reviewing what we have calculated thus far. We have started with the bosonic sector of the $\mathcal{N}=2$ supergravity and dimensionally reduced that to two dimensions for a very specific case. Our main goal is to probe the fate of F-duality in this kind of scenarios. Therefore it is quite imperative that we should keep track of the charge dependence in all our variables.  As can be seen from the two-dimensional effective action \eqref{eq:altenrate2d}, information of the initial charge configuration is embedded in the coefficient denoted by $\Theta(\cQ)$, which is fixed for a specific supergravity theory under consideration. 

As summarised briefly in \cref{sec:review}, Freudenthal duality in physics literature is defined through \cite{Borsten:2009zy,G_naydin_2010,Borsten_2013,Marrani_2013} an anti-involutive operator on the moduli space of the black hole charges as
\begin{equation}
	\hat{\cQ}^\cA=\Omega_{\cA\cB}{\dow \sqrt{|\Delta(\cQ)|}\over \dow \cQ^\cB}
\end{equation}
With $\Delta(\cQ)$ being an invariant quartic polynomial and $\Omega$ is a $2m\times 2(m$ symplectic metric ($\Omega^T=-\Omega$, $\Omega^2=-1$) with $m$ being the number of vector fields of the theory. Where the Bekenstein-Hawking(BH) entropy can be written as
\begin{equation}
	S=\pi\sqrt{|\Delta(\cQ)|}.
\end{equation}
This duality is ill-defined for \emph{small} black holes where $|\Delta(\cQ)|=0$. Contrary to the usual U-duality, where the black hole charges transform linearly, F-duality is highly non-linear as evident from the quartic polynomial defined above although both of them conserves the BH entropy. In fact following \cite{Ferrara:2011gv,Ferrara_2013,marrani2015freudenthal}, one can show that the critical points of the black hole potential are also preserved under F-duality. Freudenthal duality and its generalities also appear in various other contexts in the literature for example in the study of the multi-centered black holes\cite{Fern_ndez_Melgarejo_2014} or in the formulation of gauge theories with symplectic scalar manifolds \cite{Marrani_2013}. For other interesting avenues one may refer to \cite{Borsten_2013_2,Klemm_2017}.

Coming back to the present content, as discussed at the end of \cref{sec:review}, F-duality entails 
\begin{equation}\label{eq:F-duality}
	\begin{split}
	& \pi\hat{\cQ}^M=\Omega^{MN}\frac{\partial S(\cQ)}{\partial \cQ^N},\\
	& \hat{\hat{\cQ}}=-\cQ, \quad \text{and} \quad S(\hat{\cQ})=S(\cQ) 
	\end{split}
\end{equation}
where $\Omega^{MN}$ is $2(n+1)\times2(n+1)$ symplectic matrix with $\Omega^T=-\Omega$ and $\Omega^2=-I$. We have already derived that for our case
\begin{equation}\label{eq:finalS0}
	S_0(\cQ)={2\pi\Theta(\cQ)\over G_4}={\pi\over 3p^0}\sqrt{{4\over 3}{(3D-p^0\,q_a p^a)^3\over D}-9(p^0\, (p.q)-2D)^2}.
\end{equation}
One can now check that $S_0(\cQ)$ remains invariant as advertised under \eqref{eq:F-duality} with $S(\cQ)=S_0(\cQ)$. 
The anti-involution of this operation can be checked starting with \eqref{eq:finalS0} as the generating function for the F-dualization. In fact continual use of extremal entropy to generate F-duality indeed preserve the near-extremal entropy $S_{NE}$ as well.

The current setup enables us to ask the question about the fate of F-duality invariance beyond the extremal limit. Na\"ively one might consider that F-duality can also be generated by the full near extremal entropy (assuming $S(\cQ)= S_{NE}$). As pointed out in \eqref{eq:fullS}, the boundary condition at the overlapping region of NH and FH forces the full entropy to a non linear dependence on $\Theta(\cQ)$. One can now check that using \eqref{eq:fullS} in \eqref{eq:F-duality} by taking 
\begin{equation}
	S(\cQ)=S_{NE}={2\pi \Theta(\cQ)\over G_4}\left(1+4 \pi T\sqrt{2 \Theta(\cQ)}\right)+{3\over 2}\log\left(2\Theta(\cQ)\sqrt{2\Theta(\cQ)}{T\over G_4}\right)+{3\over 2},
\end{equation}
with
\begin{equation}
	\Theta(\cQ)={G_4\over 6p^0}\sqrt{{4\over 3}{(3D-p^0\,q_a p^a)^3\over D}-9(p^0\, (p.q)-2D)^2}.
\end{equation}
We find that in this case, F-duality mapping is not a symmetry of entropy anymore. This mapping fails to retain anti-involution property as well. Since the quantum correction appearing at the non-zero temperature \emph{diverges} at the zero temperature limit, one might consider ignoring the quantum correction discussed in \eqref{eq:fullS}, and generate the F-duality through the classical piece of the near-extremal entropy. It can be checked that even in that case, the F-duality fails to be a symmetry of the entropy. Therefore we conclude that F-duality continues to hold when the transformation is generated by the extremal entropy, though it does not hold when it is generated by the near-extremal entropy with or without the quantum correction.


\section{Conclusion}\label{concu}
In this paper, we have studied the invariance of the entropy of a near extremal, four-dimensional black hole in $\cN=2$ ungauged, supergravity under Freudenthal duality (F-duality), which is an anti involutional mapping of black hole charge vector. We have considered spherically symmetric, dyonically charged black holes in asymptotically flat space. When the black hole is extremal, supersymmetric, the entropy is F-duality invariant. As we go to the near extremal limit, the theory is well captured by JT gravity upon a dimensional reduction on $S^2$. Using the JT gravity theory as our apparatus, we find that the invariance of entropy of a near extremal black hole depends on how we define the duality itself. Originally, for a supersymmetric, extremal black hole, the duality derives a new set of charge vectors by taking derivatives of the extremal, supersymmetric entropy $S_0(\cQ)$  with respect to the corresponding charges. Considering this definition of F-duality, we find a new set of charges. We see that entropy corresponding to the new charges is the same as the near extremal entropy of original charges i.e. the entropy is F-invariant. However, when we tweak the mapping of charges, being explicit, if we define the new set of charges as the variation of the near extremal entropy $S_{NE}$, we find that in this case, near extremal entropy  is not F-invariant.  Therefore, we see that the effect of F-duality on the entropy of a near extremal black hole depends on how we define the mapping of the charge vectors.

F-duality in supergravity is broadly studied for extremal, supersymmetric black holes. In this paper, we have extended this study for a near-extremal black hole as a deviation of an extremal, supersymmetric black hole. It would be interesting to analyze the effect of Freudenthal duality for extremal, non-supersymmetric and for non-extremal black holes in both ungauged and gauged supergravity, with and without truncating the fermions. It would also be interesting to analyze the invariance of entropy under F-duality for a rotating black hole solution in supergravity. An extremal black hole in gauged supergravity is Freudental invariant \cite{Klemm:2017xxk}, thus it would be interesting to analyze the effect of Freudenthal duality for near-extremal black hole in gauged supergravity. In this work we have considered a double-extremal solution to study the near-extremal limit of F-duality where we benefit from the constant moduli fields but that also devoid us of having other parameters to control the duality transformation. The inclusion of non-constant moduli could give us a better handle on how F-duality acts away from extremality. Therefore It would be interesting to generalise our study for a near-extremal black hole without considering double extremal limit.  We keep them as future endeavours.

\bigskip\bigskip

\noindent{\bf Acknowledgments:} We would like to thank Nabamita Banerjee, Prasanta K. Tripathy, Robert de Mello Koch and Vishnu Jejjala for helpful comments on our initial draft. We also would like to thank Muktajyoti Saha for many useful discussions at many stages of the project. The work of AC and TM is supported by a Simons Foundation Grant Award ID 509116 and by the South African Research Chairs initiative of the Department of Science and Technology and the National Research Foundation. We also thank all the medical and non-medical workers who are tirelessly working and vaccinating the people in need during these troubled times.


\appendix
\section{Background Solution}\label{app:STU}
In this appendix we discuss the attractor solutions following \cite{Behrndt:1996hu}. This rather \emph{simple} case consists of three complex scalar and four gauge fields as
\begin{equation}
	x^{1}=S, \,x^{2}=T,\, x^{3}=U, \,\text{and}\, D_{abc}=\begin{cases}
		1/6 \,\forall\,\, a\neq b\neq c\\
		0 \, \text{else}
	\end{cases}.
\end{equation}
The justification of this naming comes partially from the fact that in these cases the prepotential can be written as $F=STU$. The prepotential $F$ defined earlier in \eqref{eq:theprepot} simply boils down to
\begin{equation}
	F={X^1 X^2 X^3\over X^0}.
\end{equation}

Since we are taking the double extremal black hole, we have much simply tractable set-up as the scalars $x^a=(x^a_1+i\,x^a_2)$ takes the constant value all over the space-time\footnote{In generic case, for supersymmetric attractors, one can write the moduli at any point of spacetime in terms of Harmonic functions \cite{Mandal:2016pgp}.}. $a$ denotes the number of vector multiplets, in the case of STU $a=3$. For a supersymmetric $D0-D2-D4-D6$ system, the attractor values are \cite{Shmakova:1996nz},
\begin{align}\label{eq:STUattrctr}
	x^a_1&= {3\over 2}{\tilde{x}^a\over p^0(\Delta_c \tilde{x}^c)}\left(p^0(p.q)-2D\right)+{p^a\over p^0},\\
	x^a_2&={3\over 2}{\tilde{x}^a\over (\Delta_c\tilde{x}^c)}{S\over \pi},
\end{align}
with  $q_0$ and $q_a$ denote $D0$ and $D2$ brane charges respectively, whereas $p^0$ and $p^a$ denote $D6$ and $D4$ brane charges respectively.  Where the entropy $S$ is \cite{Shmakova:1996nz},
\begin{eqnarray}
S= \frac{\pi}{3p^0}\sqrt{\frac{4}{3}(\Delta_a \tilde{x}^a)^2-9(p^0(p.q)-2D)^2}
\end{eqnarray}
and $\tilde{x}^a$ are real solution of $\Delta_a = D_{abc}\tilde{x}^b\tilde{x}^c=3D_a-p^0q_a$ with $D_a=D_{abc}\,p^bp^c$.

%

Considering an ansatz that $\tilde{x}^a=\sqrt{{3D-p^0\,q_ap^a\over D}}p^a$, we see that the entropy can be written as \cite{Mandal:2017ioi},
\begin{equation}
	S={\pi\over 3p^0}\sqrt{{4\over 3}{(3D-p^0\,q_a p^a)^3\over D}-9(p^0\, (p.q)-2D)^2}.
\end{equation}	

\bibliographystyle{jhep}
\bibliography{bib_taniya}{}

\end{document}